\documentclass[aps,prl,twocolumn,superscriptaddress,showpacs,amsmath,amssymb]{revtex4-1}

\usepackage{graphicx}
\usepackage{color}
\usepackage{textcomp}

\newcommand{\CaK}{CaKFe$_4$As$_4$}
\newcommand{\Ni}{CaK(Fe$_{0.95}$Ni$_{0.05}$)$_4$As$_4$}
\newcommand{\CaKx}{CaK(Fe$_{1-x}$Ni$_{x}$)$_4$As$_4$}
\newcommand{\BaK}{Ba$_{1-y}$K$_y$Fe$_2$As$_2$}

\begin{document}

\title{Robust $s_{\pm}$ pairing in CaK(Fe$_{1-x}$Ni$_x)$As$_4$ ($x = 0$ and $0.05$) \\from the response to electron irradiation}

\author{S. Teknowijoyo}
\affiliation{Ames Laboratory, Ames, IA 50011}
\affiliation{Department of Physics $\&$ Astronomy, Iowa State University, Ames, IA 50011}

\author{K. Cho}
\affiliation{Ames Laboratory, Ames, IA 50011}
\affiliation{Department of Physics $\&$ Astronomy, Iowa State University, Ames, IA 50011}

\author{M. Ko\'{n}czykowski}
\affiliation{Laboratoire des Solides Irradi\'{e}s, CNRS-UMR 7642 \& CEA-DSM-IRAMIS, Ecole Polytechnique, F 91128 Palaiseau cedex, France}

\author{E.~I.~Timmons}
\affiliation{Ames Laboratory, Ames, IA 50011}
\affiliation{Department of Physics $\&$ Astronomy, Iowa State University, Ames, IA 50011}

\author{M.~A.~Tanatar}
\affiliation{Ames Laboratory, Ames, IA 50011}
\affiliation{Department of Physics $\&$ Astronomy, Iowa State University, Ames, IA 50011}

\author{W.~R.~Meier}
\affiliation{Ames Laboratory, Ames, IA 50011}
\affiliation{Department of Physics $\&$ Astronomy, Iowa State University, Ames, IA 50011}

\author{M.~Xu}
\affiliation{Ames Laboratory, Ames, IA 50011}
\affiliation{Department of Physics $\&$ Astronomy, Iowa State University, Ames, IA 50011}

\author{S.~L.~Bud'ko}
\affiliation{Ames Laboratory, Ames, IA 50011}
\affiliation{Department of Physics $\&$ Astronomy, Iowa State University, Ames, IA 50011}

\author{P.~C.~Canfield}
\affiliation{Ames Laboratory, Ames, IA 50011}
\affiliation{Department of Physics $\&$ Astronomy, Iowa State University, Ames, IA 50011}

\author{R. Prozorov}
\email[Corresponding author: ]{prozorov@ameslab.gov}
\affiliation{Ames Laboratory, Ames, IA 50011}
\affiliation{Department of Physics $\&$ Astronomy, Iowa State University, Ames, IA 50011}

\date{14 February 2018}

\begin{abstract}

Controlled point-like disorder introduced by 2.5 MeV electron irradiation was used to probe the superconducting state of single crystals of \CaKx\ superconductor at $x = 0$ and 0.05 doping levels. Both compositions show an increase of the residual resistivity and a decrease of the superconducting transition temperature,  $T_c$ at the rate of $dT_c/d\rho(T_c) \approx$ 0.19 K(\textmu$\Omega$cm)$^{-1}$ for $x=0$ and 0.38 K(\textmu$\Omega$cm)$^{-1}$ for $x=\:$0.05, respectively. In Ni - doped, $x = 0.05$, compound the coexisting spin-vortex crystal (SVC) magnetic phase is suppressed at the rate of $dT_N/d\rho(T_N)\approx$ 0.16 K(\textmu$\Omega$cm)$^{-1}$. Low - temperature variation of London penetration depth is well approximated by the power law, $\Delta \lambda (T) = AT^n$ with $n\approx\,$2.5 for $x=0$ and $n\approx\,$1.9 for $x=0.05$ in the pristine state. Electron irradiation leads to the exponent $n$ increase above 2 in $x=0.05$ suggesting superconducting gap with significant anisotropy that is smeared by the disorder scattering. Detailed analysis of $\lambda (T)$ and  \(T_{c}\) evolution with disorder is consistent with two effective nodeless superconducting energy gaps due to robust s$_{\pm}$ pairing. Overall the behavior  of \CaKx\ at $x = 0$ is similar to a slightly overdoped \BaK\ at $y \approx$ 0.5 and at $x= 0.05$ to an underdoped composition at $y \approx$ 0.2.
\end{abstract}

\maketitle

\section{Introduction}

Hole-doped iron based superconductors,  Ba$_{1-y}$K$_y$Fe$_2$As$_2$ (BaK122), have a complex temperature-doping phase diagram \cite{Rotter2008}. In the parent compound, BaFe$_2$As$_2$, stripe-type magnetic order sets in at the N\'eel temperature, $T_N$, simultaneously with the tetragonal to orthorhombic structural transition at $T_s$ \cite{Canfield2010}. Upon K-doping magnetic order is suppressed and superconductivity appears at about $y\approx0.18$. Coexistence of stripe magnetic order and superconductivity in the underdoped regime ($y\leq $ 0.25) leads to substantial anisotropy of the superconducting gap \cite{KimPRB2014}, rapidly increasing in the underdoped compositions. These observations suggest that the interplay of superconductivity and magnetism may be of importance for the superconducting pairing, in line with theoretical suggestions \cite{Parker2009,IsmerEremin2010,Fernandes2010}.

The discovery of stoichiometric CaKFe$_4$As$_4$ (CaK1144) \cite{Iyo2016,Meier2016,MeierGrowth} provided unique opportunity to study effectively hole-doped system without additional scattering from chemical substitution ions. By electron count CaKFe$_4$As$_4$ corresponds to Ba$_{1-y}$K$_y$Fe$_2$As$_2$ at $y=0.5$, and, indeed, their properties are very similar \cite{Meier2016,ChoPRB2017} but with a notably lower residual resistivity in CaK1144 due to the absence of substitutional disorder. London penetration depth and STM studies of the CaKFe$_4$As$_4$ \cite{ChoPRB2017} are consistent with the two effective gaps  $\Delta_1$ and $\Delta_2$ in the range of 6-10 meV and 1-4 meV, respectively, which is close to the behavior found near $y=0.5$ Ba$_{1-y}$K$_y$Fe$_2$As$_2$ \cite{ChoSciAdv2016}.
Electron-doping of CaKFe$_4$As$_4$ with Ni and Co leads to magnetic ordering, similar to compositions with $y\leq$ 0.25, albeit of a different type, which is the spin-vortex crystal (SVC) magnetic order \cite{Meier2018}. Therefore it is of interest to probe whether this structure coexists with superconductivity differently compared to BaK122.

In this work we study the superconducting gap structure of \CaK\ in both stoichiometric and Ni-doped samples with different amount of point - like disorder. This controlled disorder is characterized by measuring normal state resistivity as described elsewhere \cite{ProzorovPRX2014}. The effect on the superconducting state is revealed by measuring the changes in $T_c$ and low-temperature variation of London penetration depth $\Delta \lambda (T)$. We find nodeless gap, along with a  rapid suppression of $T_c$ with disorder, a strong indication for $s_{\pm}$ pairing \cite{Mazin2008}. We do not find any obvious effect of the SVC type of magnetic ordering on superconductivity in Ni-doped compound.

\section{Experiment}

Single crystals of \CaKx\ ($x = 0$ and 0.05) were grown from a high temperature solution rich in Fe and As \cite{Meier2016,MeierGrowth}. The composition of Ni-doped crystals was determined by electron probe microanalysis with wavelength dispersion spectroscopy. Detailed description of crystal synthesis and  characterization are reported elsewhere \cite{Meier2016,Meier2018}. Since impurity phases are frequently observed in \CaK\ crystals, the samples for detailed  studies were selected by screening many candidates using custom made sensitive susceptometer  \cite{Tanatar2014}. Only samples with sharp transitions and no additional features on temperature scans were selected. Selected samples had typical dimensions of 0.5x0.5x0.02 mm$^3$. Four-probe electrical resistivity was measured in {\it Quantum Design} PPMS with the electrical contacts soldered with tin \cite{Tanatar2010,patent} which are preserved throughout the irradiation process, thereby removing the uncertainty of geometric factor in quantitative analysis.

In-plane penetration depth was measured using a self-oscillating tunnel-diode resonator (TDR) technique. The sample is mounted on sapphire rod using Apiezon N-grease and inserted into a 40 turns inductor without touching it. The coil is part of an LC tank circuit actively stabilized around 5 K with sub-mK accuracy. The setup is fitted inside $^3$He cryostat with sample and circuit in vacuum. The change in the resonant frequency shift due to temperature - dependent screening of small  ($\thicksim$20 mOe) excitation AC field into the sample can be calibrated by knowing the sample geometry and measuring frequency change when the sample is physically removed from the coil at low temperature. Detailed description of the technique and calibration can be found elsewhere \cite{Prozorov2000,Prozorov2006,Prozorov2011}.
Dedicated crystals for penetration depth study were also measured before and after irradiation, which are separate from the set for resistivity.

Electron irradiation was performed at SIRIUS Pelletron in Laboratoire des Solides Irradi\'es at \'Ecole Polytechnique (Palaiseau, France). The irradiation was conducted using 2.5 MeV electrons in 22 K liquid hydrogen environment to prevent immediate recombination of the vacancy-interstitial defects (Frenkel pairs) and dissipate collision energy. Upon warming to room temperature after irradiation, about 20 to 30 percent of defects are annealed as indicated by the decrease of the residual resistivity measured {\it in-situ} \cite{ProzorovPRX2014}. After this initial annealing, the defects remain relatively stable as long as the samples are kept at room temperature. We re-measured Ba$_{1-y}$K$_y$Fe$_2$As$_2$ samples after months of passive storage and did not detect any noticeable changes. The acquired irradiation dose presented in this paper is in the units of coulomb per square centimeter, where 1 C/cm$^2$ = 6.24 x 10$^{18}$ electrons/cm$^2$. The total charge of electrons penetrated through the sample was measured by the Faraday cage behind the sample stage.

\begin{figure}[tb]
        \includegraphics[width=8.6cm]{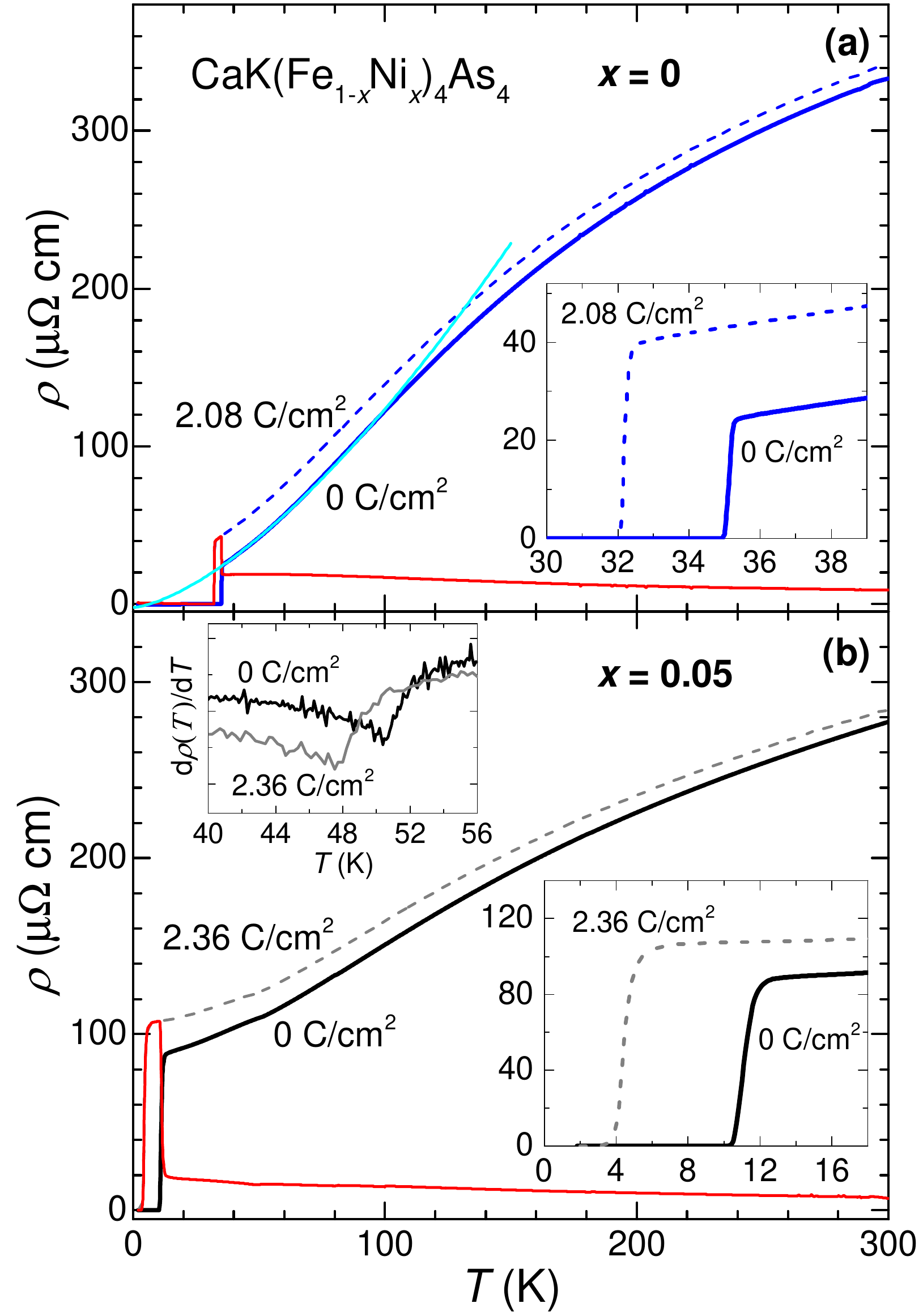}%
        \caption{\textbf{(Color online)} In-plane resistivity of stoichiometric CaKFe$_4$As$_4$ sample ($x=0$ top panel \textbf{(a)}) and of sample with Ni substitution $x=0.05$ (bottom panel \textbf{(b)}). Solid and dashed lines show resistivity of the samples before and after irradiation, with doses 2.08 C/cm$^2$ ($x=0$) and 2.38 C/cm$^2$ ($x=0.05$). Red lines show the change of resistivity between irradiated and pristine states. Cyan line in the top panel is fit of the curve in pristine $x=0$ sample to $\rho(0)+\rho_T T^{3/2}$. Right insets zoom on the superconducting transition range. Left inset in the bottom panel shows temperature-dependent resistivity derivative zooming on the features at $T_N$, suppressed upon irradiation from 50.6 K to 47.5 K.}
        \label{fig1}
\end{figure}

\section{Results and discussion}

Fig.~\ref{fig1} shows in-plane resistivity of the parent CaKFe$_4$As$_4$ (top panel) and Ni-doped  CaK(Fe$_{0.95}$Ni$_{0.05}$)$_4$As$_4$ (bottom panel) before (solid lines) and after (dashed lines) electron irradiation with 2.08 C/cm$^2$ and 2.38 C/cm$^2$, respectively. In-plane resistivity of $x = 0$ sample in pristine state shows cross-over feature at about 200~K, typical for all hole-doped compositions. Approaching $T_c$ on cooling, $\rho(T)$ shows small upward curvature, similar to Ba$_{1-y}$K$_y$Fe$_2$As$_2$ where it can be fitted with $\sim T^{3/2}$ dependence in a limited temperature range from 40 K to 60~K \cite{YLiu2014}. Similar power law fits the data well in CaKFe$_4$As$_4$ as shown in the top panel of Fig.~\ref{fig1} (cyan line). The resistivity just above the onset of resistive transition is about 12 times lower than $\rho(300K)$. The actual residual resistivity is impossible to extrapolate convincingly, since $T^{3/2}$ fit gives small negative value of $\rho(0)$ and $T_c$ is large. The resistive transition to the superconducting state at $T_c$(onset) = 35.2 K is very sharp (see lower inset in top panel of Fig.~\ref{fig1}) with width of $\Delta T_c<$ 0.5~K, reflecting good sample quality. Electron irradiation of 2.08 C/cm$^2$ leads to a vertical shift of the $\rho(T)$ curve, with red line in top panel of Fig.~\ref{fig1} showing the difference between $\rho(T)$ curves before and after irradiation. The shift is not constant, with the value just above $T_c$ is about two times higher than at room temperature, suggesting strong violation of the Matthiessen rule. The  superconducting transition remains sharp after the irradiation supporting homogeneous defect distribution. The suppression of $T_c$ with increase of residual resistivity $\rho(T_c$), happens at a rate, $dT_c$/d$\rho(T_c$) = -0.19 K(\textmu $\Omega$cm)$^{-1}$, which is remarkably close to that of slightly over-doped \BaK\ (see Fig.~\ref{fig3}(a) below for direct comparison).

\begin{figure}[tb]
        \includegraphics[width=8.6cm]{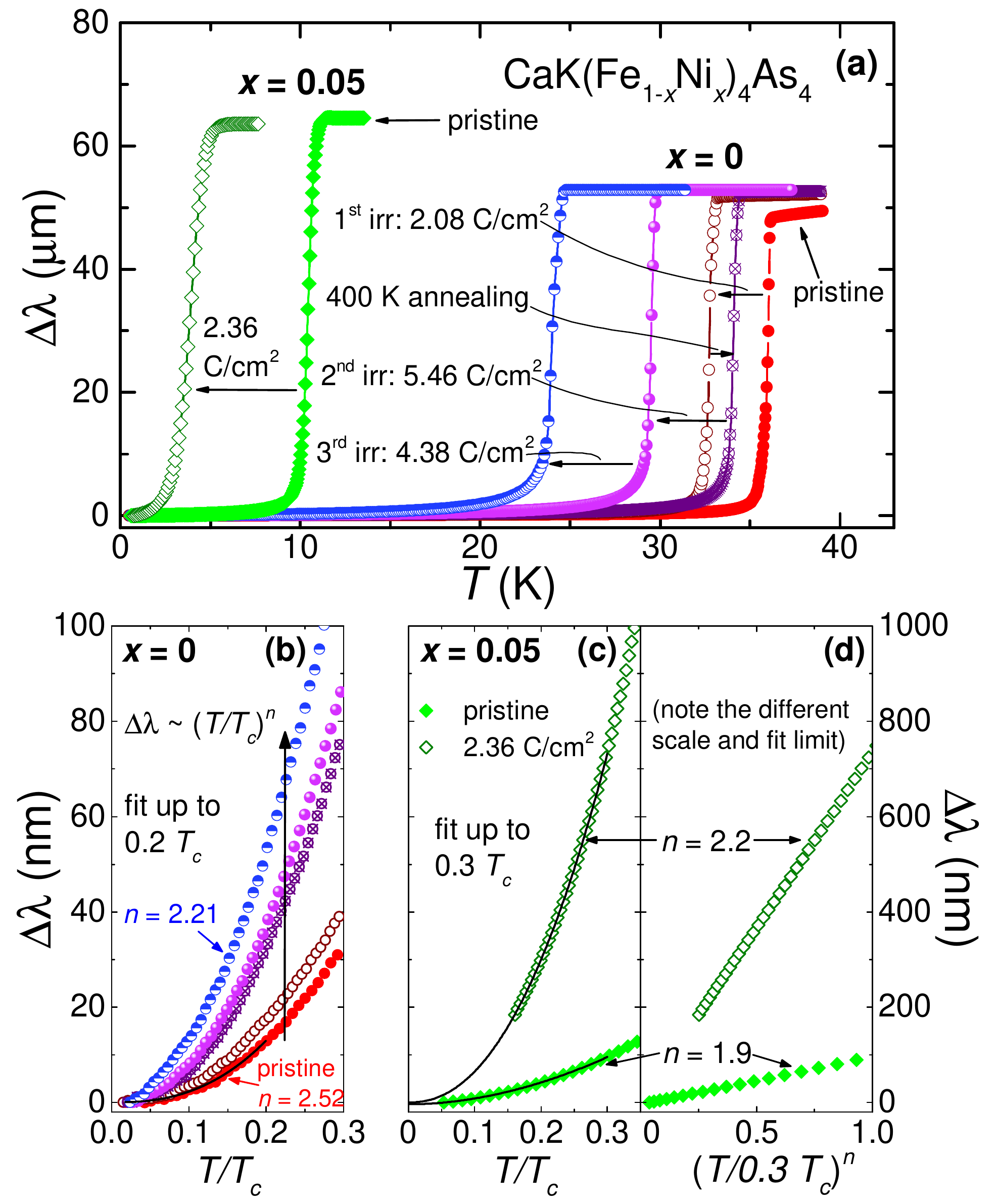}%
        \caption{\textbf{(Color online)} \textbf{(a)} Full temperature range $\Delta \lambda (T)$. For the stoichiometric ($x = 0$) sample the data were taken in a sequence of irradiation / annealing treatments as indicated in the legend. \textbf{(b)} Low temperature part of  $\Delta\lambda(T/T_c$) in $x=0$ sample. The exponent $n$ monotonically decreases with irradiation/annealing treatments in a sequence specified in top panel. Two right bottom panels show $\Delta\lambda$ in Ni-doped sample $x = 0.05$ plotted as a function of $T/T_c$ (panel \textbf{(c)}) and of $(T/T_c)^n$ (panel \textbf{(d)}).}
        \label{fig2}
\end{figure}

The electrical resistivity $\rho(T)$ of Ni - doped sample, $x = 0.05$, in the pristine condition is shown by a solid curve in the bottom panel of Fig.~\ref{fig1}. It has similar broad cross-over feature at 200~K, though it is much less pronounced due to significant increase of residual resistivity compared to pure $x=0$ compound ( $\approx$ 100 \textmu $\Omega$cm, lower inset in Fig.~\ref{fig1}(b)). An additional feature in $\rho(T)$ curves of $x=0.05$ sample can be distinguished in the temperature-dependent resistivity derivative at $\sim$50~K (top inset in Fig.~\ref{fig1}(b)), most likely due to spin-vortex crystal (dubbed ``hedgehog") magnetic ordering \cite{Meier2018}.
Electron irradiation with total dose of 2.38 C/cm$^2$ with subsequent annealing at room temperature (dashed curve in Fig.~\ref{fig1}(b)) leads to an upward shift of the $\rho(T)$ curve. Similar to the pristine sample, the shift is temperature dependent and is significantly bigger for $T<T_N\sim$ 47~K suggesting partial loss of the carrier density. The magnetic transition temperature is suppressed from 50.6~K to 47.5~K, while the superconducting transition temperature is suppressed from 10.5 to 4~K (lower inset in Fig.~\ref{fig1}(b)). The rate of $T_c$ suppression with increase of residual resistivity is substantially higher than in the sample with $x=0$ (see Fig.~\ref{fig3}(a)).
Rapid suppression of both superconducting and magnetic transition temperatures with irradiation is very similar to underdoped BaK122  \cite{ChoPRB2014} where superconductivity also coexists with magnetism, albeit with a different antiferromagnetic structure (stripe - type).

To probe the superconducting state, London penetration depth was measured before and after irradiation on the same samples. While only single irradiation dose of 2.36 C/cm$^2$ was applied to a Ni - doped ($x=0.05$) sample, multiple irradiations with intermediate annealing to 400~K were performed on pure ($x=0$) sample. Top panel of Fig.~\ref{fig2} shows total variation $\Delta \lambda (T)$ over the whole superconducting range from the base temperature of 0.4~K to above superconducting $T_c$ for all samples studied.  The superconducting transition temperature, $T_c$, was suppressed by 3.2 K from $T_c^{pristine}$(onset) = 36.1 K after first dose of 2.08 C/cm$^2$, partially recovered by 1.3 K after 400 K annealing, and then further decreased by 9.7 K after second 5.46 C/cm$^2$ and third 4.38 C/cm$^2$ irradiations, so in the end $T_c^{final}$ = 24.5 K.

\begin{figure}[tb]
        \includegraphics[width=8.6cm]{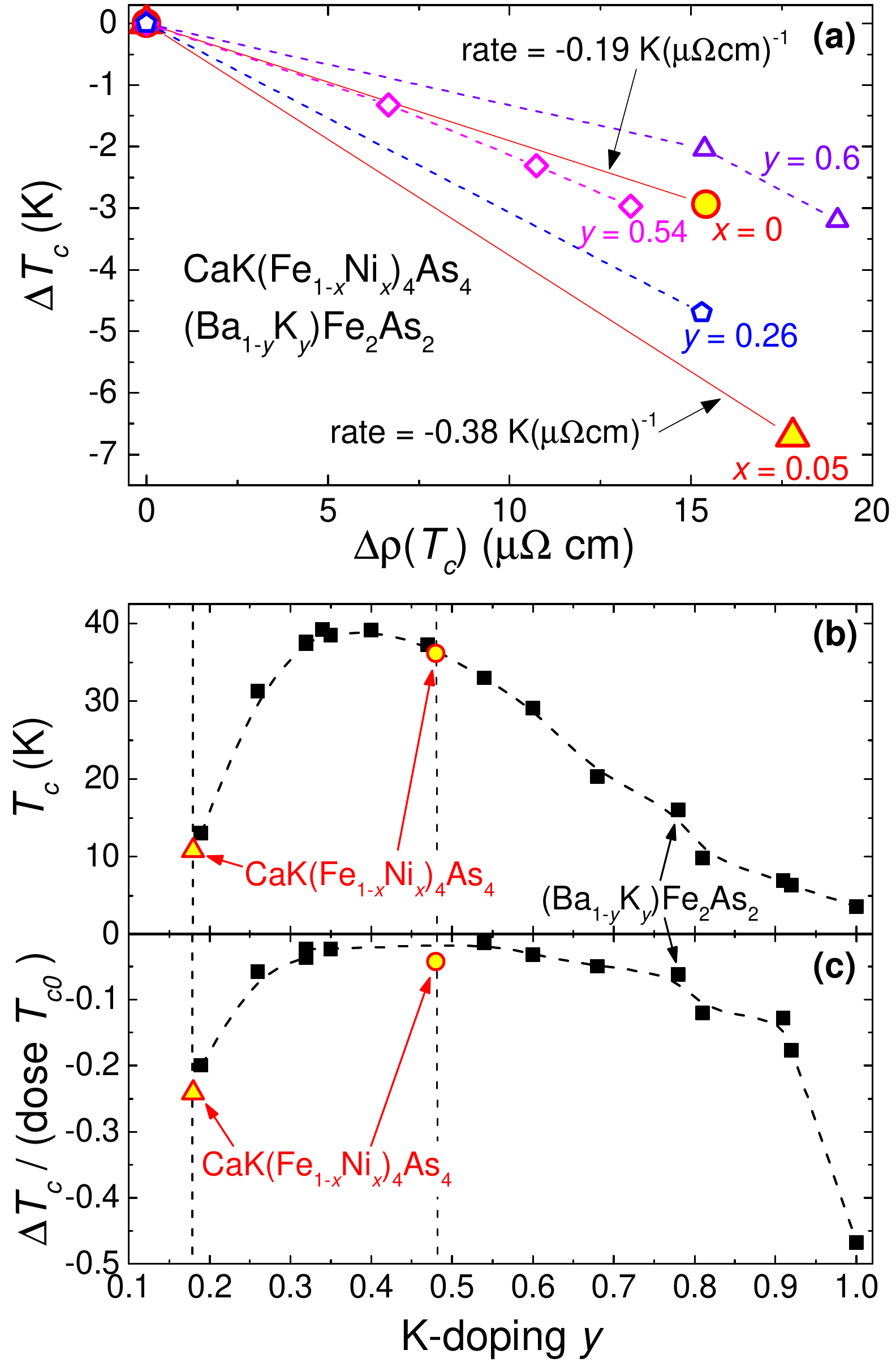}%
        \caption{\textbf{(Color online) (a)} $T_c$ suppression with irradiation parameterized via change in resistivity, $\Delta\rho(T_c)$. The rate of $T_c$ suppression in stoichiometric CaKFe$_4$As$_4$ ($x  = 0$) is similar to nearly optimally doped \BaK\ $y=0.54$ and 0.6 \cite{ChoPRB2017}. Ni - doped ($x = 0.05$) sample is close to underdoped \BaK $y=0.26$. \textbf{(b)} and \textbf{(c)} Summary of $T_c$ suppression normalized by the irradiation dose and $T_{c0}$ as a function of potassium doping $y$, where \CaK\ compounds are placed in $y= 0.18$ and 0.48 following the $T_c$ "dome" of \BaK. $T_c$ values are taken from Refs.~\cite{ChoPRB2014,ChoSciAdv2016}.}
        \label{fig3}
\end{figure}

Bottom panels of Fig.~\ref{fig2} zoom on the low-temperature part of the $\Delta \lambda (T/T_c)$ dependence. The range $T/T_c<0.3$ is considered ``low - temperatures", where the superconducting gap can be considered as practically temperature-independent and thermally excited quasiparticles reflect the gap topology in the reciprocal space. Left panel shows data for $x=0$ sample after multiple irradiations, right two panels show data for sample $x=0.05$ plotted versus reduced temperature $T/T_c$ on linear and exponent $n$ scales.
Although the annealing process slightly reversed the trend of $T_c$ evolution, the low temperature features, namely the absolute variation of the penetration depth $\Delta\lambda(T)$ = $\lambda(T) - \lambda(T_{min})$ and the exponent $n$ of the power law fit $\Delta\lambda(T)$ = A$T^n$ both exhibit monotonic behavior (see bottom left panel of Fig.~\ref{fig2}). The total variation of the London penetration depth $\Delta\lambda(T/T_c < 0.3)$ is proportional to the amount of thermally excited quasiparticles, while the exponent $n$ is a measure of pair - breaking scattering and superconducting energy gap anisotropy. Monotonic evolution of $\Delta \lambda (0.3T_c)$ as opposed to non-monotonic evolution of $T_c$ suggests that even though the defects (mostly Frenkel pairs) partially recombine during annealing to give way to higher $T_c$, there are residual low energy pair-breaking states in the underlying gap structure that are not recovered by the annealing. The exponent $n$ in $x=0$ sample gradually decreases from $n=2.5$ before irradiation to $n = 2.2$, as expected for superconductors without nodes in the gap \cite{Prozorov2011}, indicating the s$_{++}$ or s$_{\pm}$ gap structures as potential candidates. However, strong $T_c$ suppression is incompatible with s$_{++}$ pairing.

The data of Fig.~\ref{fig2} indicate that disorder is more efficient pairbreaker in the Ni - doped sample $x=0.05$ compared to the undoped counterpart $x=0$, as seen in notably higher total variation of the penetration depth between base temperature and $T/T_c=0.3$  and larger fractional $T_c$ suppression. Application of the power law fit for a temperature range from base temperature up to 0.3 $T_c$ (black lines, Fig2(c)) for $x =$ 0.05 yields $n = 1.9$, which slightly increased to 2.2 after irradiation of 2.36 C/cm$^2$. The slight increase to above $n = 2$ suggests that in pristine state, $x = 0.05$ sample is not yet in the regime where impurity scattering dominates (dirty limit) where exponent should saturate at $n=2$ with increasing disorder \cite{Korshunov2016}. Rather, in light of previous study in underdoped Ba$_{1-y}$K$_y$Fe$_2$As$_2$ in which long range magnetism gives rise to gap anisotropy \cite{KimPRB2014,Maiti2012}, the increase of exponent is consistent with the averaging of the gap structure causing the gap become more isotropic and minima to be elevated.

At this point, it is clear that the behavior of CaK(Fe$_{1-x}$Ni$_{x}$)$_4$As$_4$ with $x=0$ ($x=0.05$) projects nicely on Ba$_{1-y}$K$_y$Fe$_2$As$_2$ system at $y\approx 0.5$ ($y\approx 0.2$). Parallels drawn in the text show good matching in $T_c$ suppression rate with irradiation and the evolution of low-temperature London penetration depth variation along with the exponents of the power-law analysis. Finally, it is possible to fit \CaKx\ into \BaK\ the phase diagram, which is best illustrated in Fig.~\ref{fig3}~(b) and (c), where the comparison to \BaK\ compounds is presented by plotting $T_c$, and its sensitivity to disorder, as function of potassium doping. In Fig.~\ref{fig3} (b), following \BaK\ ``dome", stoichiometric \CaK\ can be placed in the vicinity of optimal and slightly overdoped region ($y= 0.48$), whereas electron doped \Ni\ can be positioned on the underdoped side ($y = 0.18$) where superconductivity and magnetism coexist. Figure~\ref{fig3}~(c) shows suppression of  $T_c$ normalized by the irradiation dose and $T_{c0}$ which serves as experimental measure of the sensitivity to scattering and allows comparison between different materials. Compared to BaK122, pure \CaK\ seems to be somewhat more sensitive to disorder but \Ni\ actually matches very nicely. These observations are naturally explained by the fact that pristine CaK1144 ($x=0$) is cleaner than BaK122 ($y\approx 0.5)$, so the effect of additional disorder is more pronounced. In the doped system, on the other hand, substitutional disorder is similar between two systems. It is rather remarkable that such good mapping is possible in similar, but still different and complex materials. Considered together, presented results make very strong case for robust and ubiquitous s$_{\pm}$ pairing in iron based superconductors.

\section{Conclusions}

Electron irradiation with relativistic 2.5~MeV electrons reveals rapid suppression of the superconducting  transition temperature, $T_c$, in both stoichiometric CaKFe$_4$As$_4$ and SVC antiferromagnetic Ni-doped \CaK\ with $x=0.05$, suggesting sign changing superconducting energy gap. In both cases low-temperature London penetration depth  is consistent with the nodeless
superconducting state. The two observations provide strongest support for nodeless s$_{\pm}$ pairing in these multiband superconductors. Detailed analysis shows remarkable similarity between \CaK\ compositions at $x=0$ and $x=0.05$ with hole-doped Ba$_{1-y}$K$_y$Fe$_2$As$_2$ at $y\approx 0.5$ and $y\approx 0.2$, respectively, despite the difference in the spin structure in the magnetically ordered state.

\section{Acknowledgement}
This work was supported by the U.S. Department of Energy (DOE), Office of Science, Basic Energy Sciences, Materials Science and Engineering Division. Ames Laboratory is operated for the U.S. DOE by Iowa State University under contract DE-AC02-07CH11358. WRM was supported
by the Gordon and Betty Moore Foundations EPiQS Initiative through Grant GBMF4411. We thank the SIRIUS team, B. Boizot, V. Metayer, and J. Losco, and especially O. Cavani, for running electron irradiation at Ecole Polytechnique (supported by EMIR network, proposals 11-11-0121, 13-11-0484 and 15-5788.)

\bibliographystyle{apsrev4-1}
%

\end{document}